\documentclass[sigconf,nonacm]{acmart}

\usepackage{graphicx}
\usepackage{booktabs}
\usepackage{amsmath}
\usepackage{multirow}
\usepackage{xcolor}

\begin{document}

\title{A Structured Knowledge Infrastructure for Domain-Specific Data Asset Discovery}

\author{Mengdi Chen}
\affiliation{%
  \institution{Xiaohongshu}
  \city{Shanghai}
  \country{China}
}
\email{chenmengdi@xiaohongshu.com}

\author{Yuanxin Huang}
\affiliation{%
  \institution{Xiaohongshu}
  \city{Shanghai}
  \country{China}
}
\email{huangyuanxin@xiaohongshu.com}

\author{Yulin Jiang}
\affiliation{%
  \institution{Xiaohongshu}
  \city{Shanghai}
  \country{China}
}
\email{jiangyulin@xiaohongshu.com}

\author{Wei Sun}
\affiliation{%
  \institution{Xiaohongshu}
  \city{Shanghai}
  \country{China}
}
\email{sunwei@xiaohongshu.com}

\renewcommand{\shortauthors}{M. Chen et al.}

\begin{abstract}
Enterprise data analytics agents face two structural failures: generic RAG
retrieves the wrong asset (Hit@10\,=\,19.1\%) and delivers no usage knowledge
to prevent metric misinterpretation---stemming from four root causes (C1--C4)
ranging from semantic gap and entity ambiguity to schema drift and asset-usage gap.
We present a two-layer solution deployed in the commercial advertising data warehouse at Xiaohongshu
(5,300+ Hive tables, 14 domains).
A \textbf{three-tier dual-purpose knowledge base} (179 documents, eight-section
annotation template) serves both retrieval and generation, with a closed-loop
refresh pipeline maintaining day-level freshness (one yes/no approval, 30\,s hot-reload).
The \textbf{Graph-Guided Retriever} (GGR) uses a 2,859-node knowledge graph as
a candidate gate with intent routing to deliver 71.6$\times$ token reduction.
The \textbf{Scene-Aware Ranker} (SAR) applies 19-class entity recognition and
explicit scenario annotations; negative knowledge alone contributes 25 percentage
points of Hit@10 gain.
On two 100-question benchmarks, Hit@10 rises from 19.1\% to \textbf{96.6\%}
(+77.5\,pp) and knowledge coverage from 56\% to \textbf{77\%},
at 4.84--5.33\,s end-to-end latency.
\end{abstract}

\keywords{Agentic AI, Knowledge Management, Data Asset Retrieval, Knowledge Graph, Scenario Annotation, Enterprise RAG}

\maketitle

\section{Introduction}
\label{sec:intro}

Enterprise data analytics agents face two compounding failures: finding the
right asset among thousands of overlapping candidates, and using it correctly
despite heterogeneous schemas and polysemous metrics.
Production analysis reveals four root causes:
\textbf{(C1)} high-frequency terms are non-discriminative noise for embeddings;
\textbf{(C2)} entity references require structured grounding;
\textbf{(C3)} static annotations drift as schemas evolve;
\textbf{(C4)} correct assets still produce wrong results without usage knowledge.

\paragraph{Our approach.}
We build a dual-purpose knowledge infrastructure where GGR targets C1/C2
via graph-guided narrowing and intent routing, and SAR targets C2/C4 via
entity-matched scoring and structured knowledge delivery. For C3, a closed-loop
pipeline hot-reloads LLM-drafted patches into memory within 30\,s after expert
yes/no approval (median 18\,s).
Unlike GraphRAG~\cite{edge2024local} (global synthesis) or standard RAG~\cite{lewis2020retrieval},
we use graph community clustering as a precision gate and treat explicit domain
knowledge as first-class retrieval signals. Downstream agents (SQL-writing agents,
BI report assemblers) invoke GGR and SAR as tool-style structured APIs---not as a
text-chunk retriever---receiving ranked asset IDs paired with intent-filtered
knowledge slices, and may re-query with narrowed scope during multi-step reasoning.

\paragraph{Contributions.}
\textbf{GGR}: a 2,859-node knowledge graph used as a candidate gate with
7-class intent routing and 459 synonym groups; GGR alone (asset level) reaches
Hit@10\,=\,0.73, with 71.6$\times$ token reduction vs.\ naive full-corpus RAG.
\textbf{SAR}: 19-class entity recognition and explicit scenario annotations;
\texttt{applicable} annotations alone contribute 48 pts, \texttt{not-applicable}
suppression 25 pts. The full pipeline (GGR+SAR) raises Hit@10 to 96.6\%
(+77.5\,pp over legacy).

\section{Problem Definition}
\label{sec:prelim}

We target the Xiaohongshu commercial advertising data warehouse (5,300+ Hive tables, 1,200+ BI
datasets, 14 domains) with two tasks. Prior data-discovery systems focus on
table linkage and join finding~\cite{fernandez2018aurum,zhu2019josie}, whereas
our target is semantic asset routing for downstream agent execution.
\textbf{Task~1 (Asset discovery):} given query $q$, rank assets so ground-truth
$a^*$ appears in top-$K$ (metric: Hit@10; legacy baseline: 19.1\%).
\textbf{Task~2 (Knowledge coverage):} retrieve the required usage knowledge
(table form, metric semantics, disambiguation rules) for $a^*$; we evaluate
file-level recall, leaving downstream execution (e.g., text-to-SQL
benchmarks~\cite{yu2018spider}) to future work.
Low candidate-set purity in the legacy system motivates our GGR\textrightarrow{}SAR pipeline.

\section{Knowledge Infrastructure}
\label{sec:kb}

\begin{figure*}[t]
  \centering
  \includegraphics[width=0.98\textwidth]{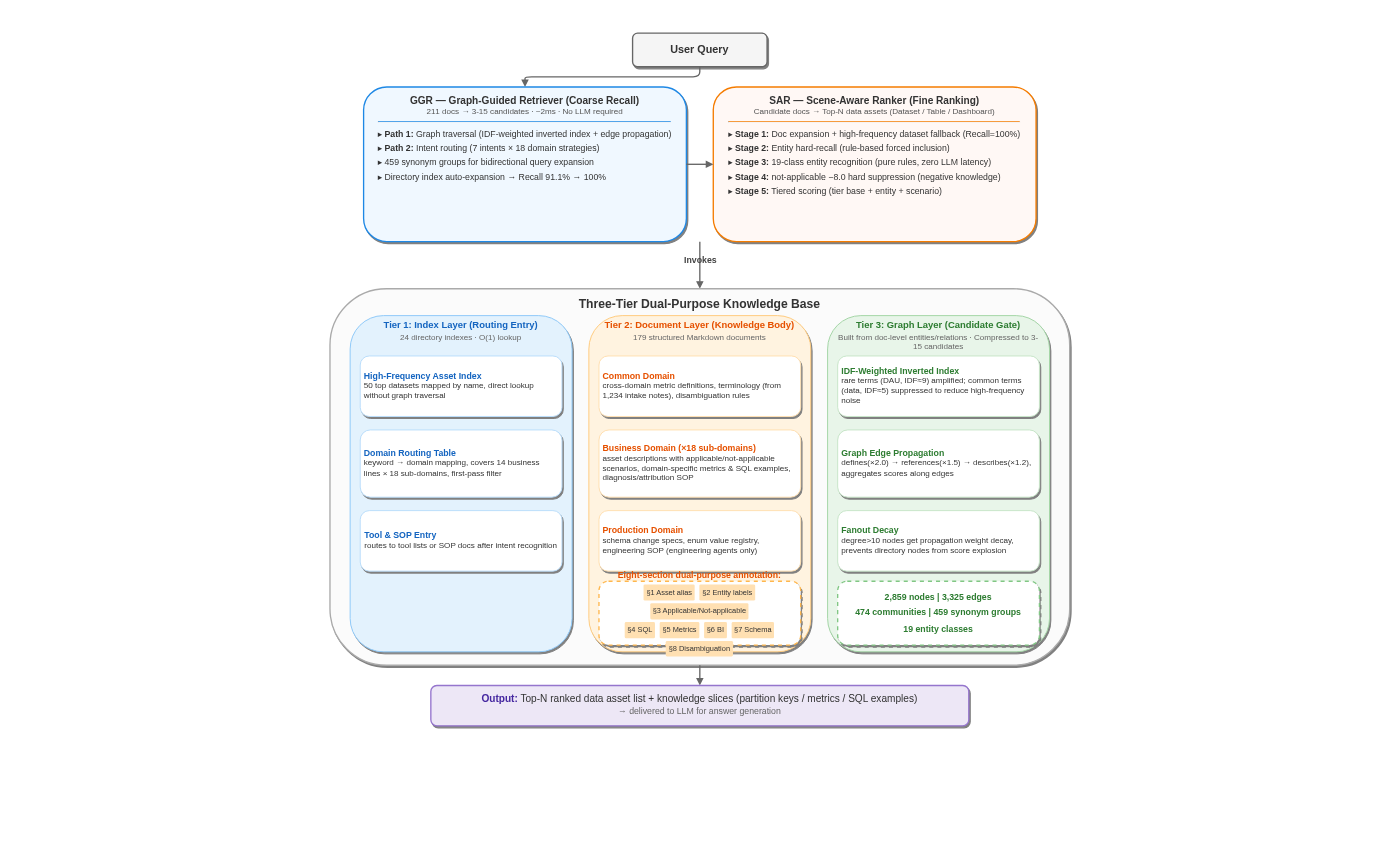}
  \caption{System overview and three-tier dual-purpose knowledge base.
    Top: GGR (coarse recall) and SAR (fine ranking) invoke the knowledge base in series.
    Tier~1: $O(1)$ keyword routing + high-frequency asset hotlist.
    Tier~2: 179 structured documents with eight-section annotation.
    Tier~3: 2,859-node knowledge graph as candidate gate (474 communities).}
  \label{fig:kb}
\end{figure*}

Our knowledge infrastructure is structured into three tiers (Fig.~\ref{fig:kb}) to balance routing speed with semantic richness.

\textbf{Tier~1} (routing table + hotlist): $O(1)$ keyword routing to the 50 most-queried assets or tool/SOP documents.
\textbf{Tier~2}: 179 Markdown documents (Common / Domain / Production) each with an eight-section annotation---\S~1--2 (aliases, entity labels) retrieval-oriented; \S~3 (applicable/not-applicable scenarios) for retrieval and generation; \S~4--8 (SQL, metrics, BI, schema, disambiguation) as 480-token intent-filtered slices (vs.\ 2,400-token full documents).
\textbf{Tier~3}: knowledge graph~\cite{graphify} with 474 Louvain communities~\cite{blondel2008louvain}, each mapping to 3--15 candidates per query.
For C3, LLM-drafted patches hot-reload within 30\,s~(P95; median 18\,s, $\sim$45/month) after expert yes/no approval, mitigating annotation drift through continual maintenance~\cite{de2021continual}.

\section{Graph-Guided Retriever (GGR)}
\label{sec:ggr}

\begin{figure*}[t]
  \centering
  \includegraphics[width=0.98\textwidth]{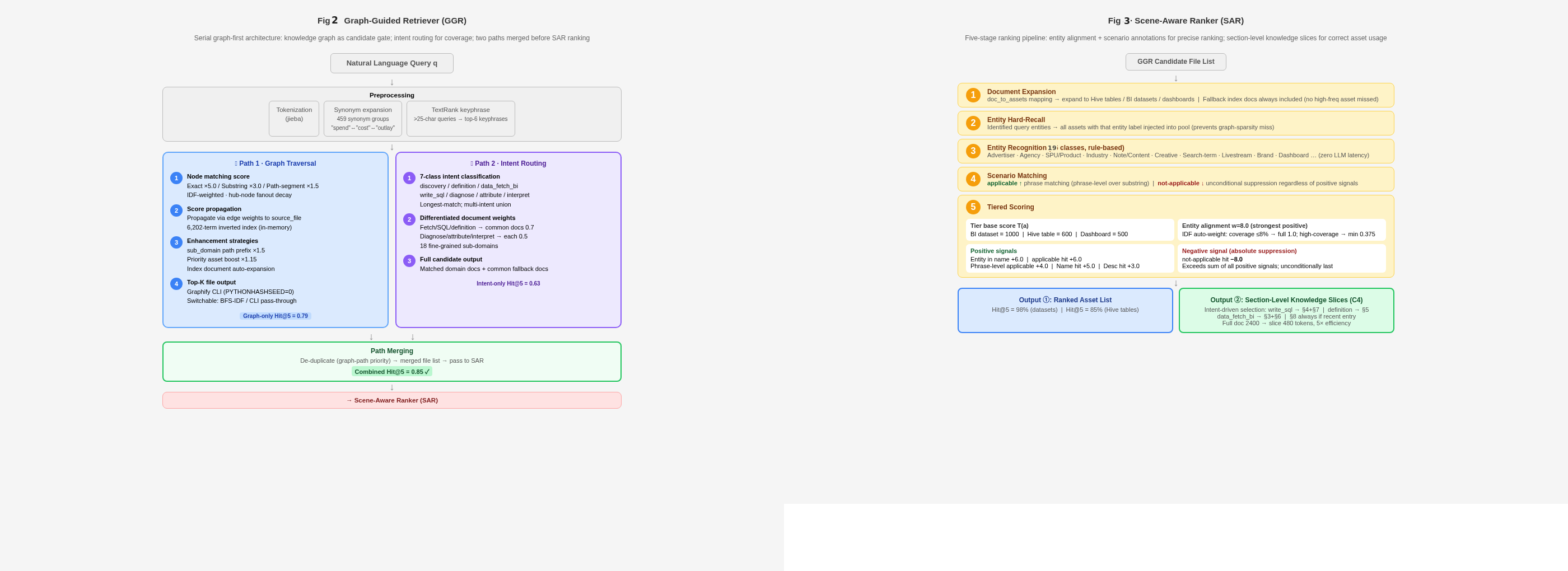}
  \caption{Overview of retrieval and ranking pipeline. Left (GGR): dual-path retrieval
    merging graph traversal (Path~1) and intent routing (Path~2). Right (SAR):
    five-stage ranking pipeline producing ranked assets and knowledge slices.}
  \label{fig:ggr}\label{fig:sar}
\end{figure*}

GGR (Fig.~\ref{fig:ggr}, left) treats the knowledge graph as a \emph{candidate gate}: graph traversal
narrows the search space before any vector comparison, constraining retrieval
to semantically relevant communities.
The GGR index covers 211 documents (179 Tier-2 content documents plus 32 directory-style index documents—cross-tier aggregate files that each link to multiple content documents).
\subsection{Knowledge Graph Construction}

We build the graph via deterministic AST parsing (confidence~1.0) and LLM
relation extraction (confidence~0.6--0.9); the confidence scoring helps mitigate
LLM hallucination in graph edges~\cite{ji2022survey}.
The graph has 2,859 nodes (1,874 concept, 526 code, 440 document, 19 rationale) and 3,325 edges across three relation types (Doc$\leftrightarrow$Concept 2,712; Concept$\leftrightarrow$Concept 562; Code/Rationale 51).

\subsection{Dual-Path Architecture}

The two paths run in parallel and complement each other:
\textbf{Path~1 (Graph traversal)} provides broad semantic coverage: it expands
query tokens via 459 synonym groups, matches at three precision levels
(exact/substring/path-segment), and propagates scores through a 6,202-term
inverted index (built from graph node labels and synonym expansions) with
IDF weighting and hub-node fanout decay. Caller-specified domains receive
a path-prefix boost; high-frequency core datasets are pinned as fallbacks
to prevent them from being ranked out by rare-term bias.
\textbf{Path~2 (Intent routing)} provides high-precision directional recall:
it classifies the query into 7 intent classes
(\texttt{discovery}, \texttt{definition}, \texttt{data\_fetch\_bi},
\texttt{write\_sql}, \texttt{diagnose}, \texttt{attribute}, \texttt{interpret})
and maps each intent directly to its specialist documents.
The two paths merge with graph priority. On the same 100-question Task~1
benchmark at \emph{document level}: graph-only Hit@10\,=\,0.79,
intent-only\,=\,0.63, combined\,=\,0.85 (asset-level in Table~\ref{tab:ablation}).
By first narrowing thousands of documents to 3--15 community candidates,
GGR lets downstream RAG operate in a low-noise space without depending on a large LLM.
GGR alone reaches asset-level Hit@10\,=\,0.73\,(est.), lower than the
document-level 0.85 due to the stricter one-to-many doc\textrightarrow{}asset
mapping; SAR then lifts asset-level to 0.966.

\paragraph{Implementation.} The intent classifier is a keyword-rule matcher
over 183 domain-specific trigger terms (7 classes) with LLM fallback for
unmatched queries---avoiding fine-tuning cost while maintaining single-digit-ms
latency. Entity recognition (19 classes, Section~\ref{sec:sar}) uses rule-based
matching against curated vocabularies (2--40 aliases per entity), tuned on
production query logs. The 459 synonym groups were semi-automatically
expanded from 120 seed terms via embedding-based clustering plus expert review;
the 6,202-term inverted index is auto-rebuilt on every KB refresh.

\section{Scene-Aware Ranker (SAR)}
\label{sec:sar}

SAR (Fig.~\ref{fig:sar}, right) elevates ranking from keyword matching to
\emph{semantic-scene alignment}---the joint optimization of (i)~entity-level
alignment between query business terms and dataset entity tags, and
(ii)~scenario-level alignment between query intent and dataset
\texttt{applicable}/\texttt{not-applicable} descriptions.

\subsection{Five-Stage Pipeline}

SAR processes candidates in five stages: (1) expand document hits to individual assets via a precomputed
\texttt{doc}$\to$\texttt{asset} mapping with high-frequency fallback indexes; (2) hard-recall
entity-labeled assets to counter graph sparsity; (3) identify 19 entity classes via rule-based matching (zero LLM latency)---spanning advertiser-centric (advertiser, agency, brand, targeting), content (creative, note, note keyword, commercial note type, IP project), data (SPU, third-party platform, reputation pass, strategy) and query dimensions (search term, upstream/downstream term, back-search term, region, user, enterprise account)---aligning query business terms with dataset entity tags;
(4) match \texttt{applicable}/\texttt{not-applicable} scenario phrases;
(5) compute tiered scores.

\paragraph{Stage~5 (Tiered scoring).}
A resource-type tier $T(a)\in\{1000,600,500\}$ (BI dataset / dashboard / Hive table)
enforces a hard hierarchy. Positive signals include IDF-weighted entity alignment
(+8.0, dominant), asset-name match ($+6.0$), and \texttt{applicable}
annotation match ($+6.0$, with an extra $+4.0$ for full-phrase matches);
the single negative signal \texttt{not-applicable} ($-8.0$) suppresses false
positives regardless of other signals. IDF weighting $\phi(e)=\max(0.375,\,1{-}0.8C(e))$,
where $C(e)$ is the corpus-level document frequency of entity $e$ across all 179
files, discounts ubiquitous terms (e.g., ``advertising'', $C\!=\!86\%\to0.375$)
and preserves rare entities (brand, $C\!=\!0.3\%\to1.0$).
Queries involving organizational hierarchy (sales / operations / industry / track)
trigger a targeted boost on wide-table assets---addressing high-frequency
structural queries that keyword matching alone struggles with.

After ranking, SAR selects intent-driven slices from \S~4--8
(\texttt{write\_sql}$\to$\S~4+\S~7; \texttt{definition}$\to$\S~5;
\texttt{data\_fetch\_bi}$\to$\S~3+\S~6) and delivers them to the LLM as in-context knowledge~\cite{shi2023rethinking}.

\section{Evaluation}
\label{sec:eval}

\subsection{Setup}

\textbf{Benchmark.} Two separate 100-question production benchmarks: Benchmark-A for Task~1 (asset discovery) and Benchmark-B for Task~2 (knowledge coverage), each sampled independently from production query logs, evaluated against the live system (v9) vs.\ the legacy production system.
The legacy system performs BM25-style full-corpus retrieval over 179 documents without graph/entity signals, serving as the primary baseline; embedding-based comparisons require domain-specific fine-tuning in this 14-domain warehouse and are left for future work.
\textbf{Task~1} (Asset Discovery): ground truth is the canonical asset over 5,300+ Hive tables and 1,200+ BI datasets across 14 domains. \textbf{Task~2} (Knowledge Coverage): each question is annotated with 1--2 must-read files; coverage = recall of ground-truth files (before slicing).

\subsection{Main Results}

\begin{table*}[t]
\centering
\caption{Main results on two 100-question production benchmarks.
  \textbf{Top}: Task~1 asset discovery (Hit@K). Legacy Hit@5\,=\,Hit@10 because
  the system returns exactly 10 candidates.
  \textbf{Bottom}: Task~2 knowledge coverage (file-level recall before slicing).}
\small
\begin{tabular}{lrrrrc}
\toprule
\multicolumn{6}{l}{\textit{Task~1: Asset Discovery (Hit@K)}} \\
\midrule
Method & Hit@1 & Hit@3 & Hit@5 & Hit@10 & Latency \\
\midrule
Legacy system  & 0.124 & 0.169 & 0.191 & 0.191 & 2.97\,s \\
\textbf{Ours (v9, online)} & \textbf{0.494} & \textbf{0.888} & \textbf{0.933} & \textbf{0.966} & \textbf{4.84\,s} \\
\midrule
\multicolumn{6}{l}{\textit{Task~2: Knowledge Coverage}} \\
\midrule
Method & Coverage & Files/Query & Latency & & \\
\midrule
Legacy system & 56.0\% & 4.5 & 4.03\,s & & \\
\textbf{Ours (v9, online)} & \textbf{77.0\%} & \textbf{6.5} & \textbf{5.33\,s} & & \\
\bottomrule
\end{tabular}
\label{tab:main}
\end{table*}

As shown in Table~\ref{tab:main}, our system achieves Hit@10\,=\,96.6\% vs.\ 19.1\% (legacy), \textbf{+77.5\,pp},
Hit@1 from 12.4\% to 49.4\%, and knowledge coverage from 56\% to \textbf{77\%} (+21\,pts).
On a standalone SAR benchmark (100 questions, independently constructed from asset-ranking logs, separate from Benchmark-A/B), SAR alone achieves
Hit@5\,=\,98\%, Recall\,=\,100\%, MRR\,=\,0.803.
Despite retrieving more files per query (6.5 vs.\ 4.5), token consumption is
71.6$\times$ lower vs.\ naive full-corpus RAG, because our system delivers
480-token intent-filtered slices rather than full 2,400-token documents
($(179 \text{ docs}\times2{,}400)\div(12.5\times480)$, where 12.5 is the measured average number of GGR candidate documents per query).
Latency overhead (+1.87\,s Task~1, +1.30\,s Task~2) stems from graph traversal and entity recognition; both remain within SLA and are negligible relative to analysts' prior manual search of several minutes per query.

\subsection{Ablation Study}

\begin{table}[h!]
\centering
\caption{Ablation (Hit@10). Each $\Delta$ is marginal contribution of the removed component.}
\small
\begin{tabular}{lrc}
\toprule
Configuration & Hit@10 & $\Delta$Hit@10 \\
\midrule
\textbf{Ours (v9, online)}           & \textbf{0.966} & --- \\
\quad w/o applicable annotation      & 0.486 & $-0.480$ \\
\quad w/o graph path (intent only)   & 0.606 & $-0.360$ \\
\quad w/o not-applicable annotation  & 0.716 & $-0.250$ \\
\quad w/o entity recognition         & 0.786 & $-0.180$ \\
\quad w/o IDF entity weighting       & 0.846 & $-0.120$ \\
\bottomrule
\end{tabular}
\label{tab:ablation}
\end{table}

Table~\ref{tab:ablation} shows \texttt{applicable} annotations contribute the most (--48\,pp);
graph path --36\,pp; \texttt{not-applicable} suppression --25\,pp (negative knowledge
matters as much as positive); entity recognition --18\,pp and IDF weighting --12\,pp confirm the value of fine-grained entity modeling. All ablations (0.486--0.846) far exceed legacy (0.191); $\Delta$ values are single-factor removals and not additive due to feature interactions.

\section{Deployment Notes}
\label{sec:lessons}

Two engineering decisions shaped the deployed system. First, C3 patches use
expert yes/no approval (median 18\,s) rather than fully automated LLM merges:
in early pilots, unsupervised auto-merges surfaced schema-drift-induced regressions
that were expensive to roll back; a lightweight human gate at 45 patches/month
is affordable and eliminates this failure mode. Second, GGR and SAR are
deliberately staged (coarse\textrightarrow{}fine) rather than fused in parallel:
in our warehouse, parallel fusion left too much noise in the candidate set for the
downstream ranker to filter effectively, whereas the staged pipeline lets SAR
operate on an already high-precision candidate pool.

\section{Conclusion}
\label{sec:conclusion}

We present a knowledge infrastructure that serves as the retrieval and knowledge layer of a production agentic analytics pipeline: agents invoke GGR and SAR to ground every query in structured domain knowledge before SQL generation or BI report assembly---directly realising the ``Agentic AI Goes Live'' agenda. Deployed in the Xiaohongshu commercial advertising data warehouse, GGR and SAR achieve 96.6\% Hit@10 (+77.5\,pp) and 77\% knowledge coverage (up from 56\%, +21\,pp), while reducing token footprint by 71.6$\times$ vs.\ naive full-corpus RAG, at 4.84--5.33\,s end-to-end latency. Our results show that in specialized domains, explicit domain knowledge---graph structure, intent routing, and scenario annotations---dominates generic embeddings. The eight-section template is domain-agnostic; adapting to a new domain requires document authoring ($\sim$2\,h/doc) with automated graph construction. Future work: cross-domain transfer and closed-loop agentic execution.

\bibliographystyle{ACM-Reference-Format}
\bibliography{references}

\end{document}